\documentclass[runningheads]{llncs}
\usepackage[T1]{fontenc}
\usepackage{bbding}
\usepackage{graphicx}
\usepackage{svg}

\usepackage{xcolor}

\begin{document}
\title{Adaptive Security in 6G for Sustainable Healthcare}

\author{Ijaz Ahmad \Envelope \inst{1}\orcidID{0000-0002-6152-8947} \and
Ijaz Ahmad\inst{2}\orcidID{0000-0003-1101-8698}\and
Erkki Harjula\inst{1}\orcidID{0000-0001-5331-209X}}

\authorrunning{I. Ahmad et al.}
\institute{Centre for Wireless Communications, University of Oulu, Finland\\
\email{\{ahmad.ijaz,erkki.harjula\}@oulu.fi} \and
VTT Technical Research Centre of Finland, Finland\\
\email{ijaz.ahmad@vtt.fi}}

\maketitle

\begin{abstract}

6G will fulfill the requirements of future digital healthcare systems through emerging decentralized computing and secure communications technologies. Digital healthcare solutions employ numerous low-power and resource-constrained connected things, such as Internet of Medical Things (IoMT). However, the current digital healthcare solutions will face two major challenges. First, the proposed solutions are based on the traditional IoT-Cloud model that will experience latency and reliability challenges to meet the expectations and requirements of digital healthcare, while potentially inflicting heavy network load. Second, the existing digital healthcare solutions will face security challenges due to the inherent limitations of IoMT caused by the lack of resources for proper security in those devices. Therefore, in this research we present a decentralized adaptive security architecture for the successful deployment of digital healthcare. The proposed architecture leverages the edge-cloud continuum to meet the performance, efficiency, and reliability requirements. It can adapt the security solution at run-time to meet the limited capacity of IoMT devices without compromising the security of critical data.  Finally, the research outlines comprehensive methodologies for validating the proposed security architecture. 

\keywords{6G  \and Adaptive Security \and Healthcare \and Edge \and Internet of Things}
\end{abstract}

\section{Introduction}

6G is the next-generation wireless communication technology with promising unprecedented speed and low latency~\cite{iturecommend}. In addition to evolutionary improvements in performance to its predecessors, 6G will also include novel innovations such as convergence of computing, communication, and sensing ~\cite{adapsec}. It also supports better integration with AI/ML by enabling the extension of edge computing to the local level (further closer to UEs), and traditional mobile operator and customer organizations’ communication infrastructures. It also facilitates the use of renewable energy sources, within and outside the network infrastructure, as well as it is envisaged to support quantum computing/communication requirements. IMT-2030 recommendations~\cite{iturecommend} comprise of novel capabilities such as artificial intelligence (AI) and sensing, ubiquitous connectivity, and integrated sensing which were not described in IMT-2020 (5G).

Digital healthcare applications, including decentralized clinical trials, mobile imaging, computer-assisted surgery, emergency response and telemedicine, among many others, have considerable low-latency and stringent privacy requirements governed by regulatory authorities. In a digital healthcare solution, numerous connected things (IoMT) help in monitoring, processing, storing, and transmitting the patient’s data. There is a need to protect the patient's data from potential exposure in full compliance with stricter regulatory and legislative requirements for privacy and security~\cite{iohtsec}. Conventional IoT-cloud setup faces the challenges of latency and exposure of sensitive data to a wider audience consisting of potentially hostile nodes.

The potential to process data at the network edge near the data source enables lower communication costs, reduced latency, enhanced efficiency, and increased system capacity for compute-intensive healthcare applications. In general, IoMT devices deployed in healthcare scenarios are resource-constrained and  unable to sustain conventional heavyweight security solutions. Hence, there is a need to provide a context-aware and resource-efficient adaptable security architecture protecting sensitive information in dynamic network conditions, available resources, and evolving threat landscape. In this research, we propose an adaptive security architecture that leverages the potential emerging technologies such as distributed ledger technologies, zero-trust~\cite{zta6g} security, differential privacy, and federated learning in a resource-constrained environment.

The rest of the paper is organized as follows: Section 2 describes the background and related work, and Section 3 presents our position statement. Section 4 provides robust evidence and reasoning to support our position and ensures our arguments are grounded in fact and logic. Section 5 comprises counterarguments and rebuttals, addressing the opposing perspectives hands-on, and Section 6 reflects on the discussions presented in the article. 
\section{Background \& Related work}
\subsection{Healthcare in 6G}
The advancement of wireless communication networks has had a profound effect on the healthcare industry. The shift from 4G to 5G initiated a new era in digital healthcare~\cite{shealth6g}, facilitating the emergence of the Intelligent Internet of Healthcare Things (IIoHT) or Internet of Medical Things (IoMT). This technology transition has created new possibilities in the healthcare sector, enhancing the efficiency, resilience, sustainability, affordability, and widespread availability of services. The expected shift to 6G is poised to profoundly transform healthcare~\cite{shealth6g},~\cite{edgeforhealth}. 6G will be a comprehensive system that will combine sensors, mobile communications, and processing capacity to effectively link virtual and actual environments. In healthcare scenarios, this implies that, e.g., the vital signs of patients may be detected, analyzed, and transferred immediately over the 6G network. 6G is anticipated to enable physicians and nurses to collaborate in new ways by using advanced network features, such as augmented reality (AR) apps or telemedicine.

\subsection{Security challenges in IoMT}
The start of the 21st century witnessed a broad shift of enterprises from local data centers to cloud infrastructure considering its flexibility and scalability, resulting in the expansion of the IoT ecosystem~\cite{eccmotivation}. Lack of standardized protocols, limited security controls, and a focus on functionality over security has led to a wide range of vulnerabilities exploited in critical infamous security incidents notably Marai botnet~\cite{marai}, Stuxnet (2010), and Vastaamo health data breach~\cite{vaastabreach}. Moreover, the use of 6G technology in the healthcare sector poses security and privacy concerns, particularly due to the sensitive nature of health data~\cite{iohtsec}. Furthermore, hostile attackers could compromise implantable medical devices (IMDs)~\cite{iotsec} such as diabetic insulin pumps, internal heart defibrillators, and pacemakers, causing not only economic losses but endangering human lives. It is essential to tackle these challenges to effectively integrate healthcare based on 6G technology. Notwithstanding these barriers, the potential advantages of 6G in healthcare are vast, and it is anticipated that the technology will enable healthcare to be heavily reliant on AI and 6G communication technology.

\subsection{Adaptive security}
Adaptive security indicates that security solutions may be adjusted on the fly to accommodate changing network conditions and evolving threats. Adaptive security helps protect healthcare services and data by encompassing mechanisms for prediction, detection, mitigation, and prevention of threats to privacy and security~\cite{spvision6g}. Existing approaches such as risk-based ~\cite{riskadapt}, Game-based \cite{gameadapt}, Requirements-driven \cite{reqadapt}, and event-driven \cite{eventadapt} adaptive security models will be evaluated for edge-cloud continuum considering their applicability and adaptability.

\subsection{Resource efficient healthcare solutions in 6G}

Sustainability is set as one of the new capabilities for IMT-2030 and security, privacy, and resilience as enhanced capabilities over the existing 5G (IMT-2020)~\cite{iturecommend}. Enhanced mMTC and uRLLC from 5G will pave the way for ubiquitous connectivity as an emerging use case in 6G that will call for robust and sustainable~\cite{adaptivecyber} security, privacy, and trust measures~\cite{adapsec}. Conventional security measures and strategies use pre-set, manual methods to reduce risks and keep things safe in certain situations ~\cite{adaptivefeasibility} at the cost of additional energy consumption resulting in larger energy footprints. The proposed adaptive security architecture will consider the available resources especially remaining energy while providing security in a healthcare scenario. 

\section{Position Statement}
\begin{figure}[htb!]
  \centering
  \includegraphics[width=\linewidth]{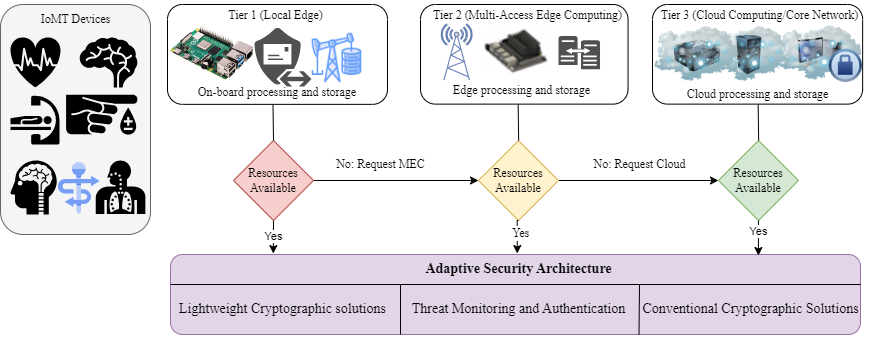}
  \caption{Adaptive Security at Edge}
  \label{fig:adaptsec}
\end{figure}

The development of the 6G and edge-cloud continuum will extend the widespread adoption of IoT in many delay-sensitive scenarios in healthcare~\cite{edgecritical}. Conventional stringent security mechanisms are resource-intensive and not flexible across the edge-cloud continuum. Hence, there is a need for context-aware adaptive security architecture considering the remaining energy, network dynamics, and evolving threat landscape. Adaptive security is a promising approach to address the rising security challenges by distributing security capabilities across the network, from local edge to MEC and ultimately to the cloud as illustrated in Figure \ref{fig:adaptsec}. The IoMT devices and the underlying network are monitored for security breaches from MEC (tier-2) in addition to computing delay-sensitive tasks that are too heavy for processing at the local edge. Local edge (tier-1) hosts lightweight cryptographic solutions with strict latency requirements, while delay-tolerant and computationally demanding security tasks are assigned to the cloud layer (tier-3). This enables adapting the security measures to the changing network conditions and the healthcare use-case requirements at run-time with no human intervention.

\section{Evidence and Reasoning}
\subsection{Performance and efficiency}
In many digital healthcare scenarios, the key requirements include the ability to communicate in real-time with no observable latency. One such example is emergency patient monitoring to communicate the patient's vitals from the incident site to the hospital. In such scenarios, secure communication is among the key features, in addition to performance and reliability. Adaptive security ensures that the communication is protected from potential disclosure while considering the available resources such as energy, processing, and network conditions.
\begin{figure}[htb!]
  \centering
  \includegraphics[width=0.9\linewidth]{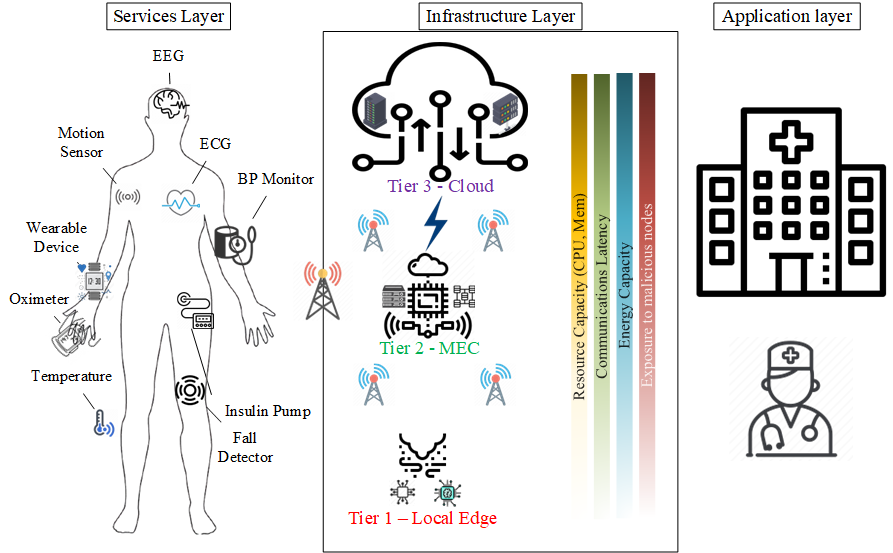}
  \caption{Secure Digital Healthcare Architecture}
  \label{fig:healthcare}
\end{figure}
As illustrated in Figure \ref{fig:healthcare}, the capacity of computational resources and energy reduces while moving from Tier-3 (cloud) towards Tier-1 (local edge). Fortunately, the exposure to security threats (from potential hostile/malicious nodes) reduces in line with the available computational and energy resources. From a perspective of security and privacy, the computations performed in close proximity are preferred, whenever feasible, as the data do not need to be transmitted over the air for a longer distance and these are therefore less exposed to attackers.
\subsection{Heterogeneity \& Diversity}
Heterogeneity refers to the variety and diversity of standards and technology included in medical devices. 6G networks are anticipated to facilitate a substantial volume of interconnected devices and transfer data of high intensity~\cite{diverseiomt},~\cite{cardiohybrid}. It is essential to tackle the diversity and range of IoMT devices to guarantee the security of 6G-enabled healthcare services and data. Traditional security solutions are inadequate in a diverse IoMT environment with a dynamic and evolving technological and threat landscape. The envisioned heterogeneity feature of future 6G networks increases the attack surface and may result in complex security challenges~\cite{tspchallenge}.

\subsection{Technological Advancements}
The IMT-2030 recommendations~\cite{iturecommend} have identified distributed ledger technologies (DLT such as blockchain), differential privacy, and federated learning as potential security technologies for achieving security and resilience in future 6G networks. Utilizing the potential of blockchain technology for authentication and securing healthcare data in a distributed manner with no or reduced reliance on centralized servers will improve the efficiency of the network. Similarly, with the breakthrough of Artificial Intelligence and Machine Learning based solutions can be developed for detecting and responding to threats efficiently in real-time.

\subsection{Lightweight Cryptography}
Security specialists at the National Institute of Standards and Technology (NIST) have declared a champion in their quest to identify a robust defender for data produced by resource-constrained devices. The triumphant contender, a set of cryptographic algorithms known as ASCON~\cite{ascon12}, has been recognized as NIST’s standard for lightweight cryptography since February 2023. This helps in providing sustainable and affordable security in IoT-based resource-constrained devices whereby conventional heavyweight security solutions are considered as barriers to performance and survival.

\subsection{Limited exposure to Malicious Agents}
By employing effective and adaptive security solutions across the edge-to-cloud continuum ensures the protection of sensitive data closer to the origin and thus reduces the geographical area of exposure to potential malicious agents. The widespread popularity and adoption of IoMT devices are continuously contributing to novel security threats and their complexities are increasing over time. The vulnerabilities in communication networks can be broken down into the core technologies and potential solutions can be looked into to achieve overall security~\cite{secsurvy}. Adaptive security ensures timely protection of data in a resource-constrained IoMT environment.

\section{Counterarguments and Rebuttals}
\subsection{Complexity and Cost}
\textit{\textbf{Implementing adaptive security in 6G for sustainable healthcare could increase complexity and cost.}} Adaptive security is the ability to adapt to threats in real time and can potentially save costs associated with data breaches and system downtime. Recent research indicates that the long-term advantages of incorporating adaptive security in healthcare offered by 6G technology surpass these early obstacles of increased complexity and cost. The traditional approach of using a single security strategy is no longer suitable for 6G networks~\cite{adapsec}. This is because there is a greater variation in device capabilities, energy conditions, service features, threats \& vulnerabilities, and other attributes that change over time. Therefore, the selection and configuration of security strategies for 6G networks should be optimized flexibly and responsively. By using this adaptable strategy, it is possible to optimize resource use, which may result in a decrease in overall expenses over time.

\subsection{Technology Maturity}
\textit{\textbf{6G technology is still in its infancy and it is too early to discuss its security aspects.}} IMT-2030 recommendation~\cite{iturecommend} describes 6G to be secure by design with the ability to continue operating in the presence of malicious agents and be able to recover from disruptive events. Recent research ~\cite{roadmap6g} suggests that the discussion and preparation for security concerns of 6G technology should begin without delay, even before the technology reaches full maturity. Instead, security issues should be included in the development process from the beginning. Implementing this strategy may facilitate the early detection and resolution of possible security vulnerabilities, resulting in the development of more resilient and secure 6G networks.

\subsection{Privacy Concerns}
\textit{\textbf{Patient privacy is impacted by the increased connectivity and data sharing in 6G-enabled healthcare.}} 6G technology promises to make healthcare more efficient, robust, sustainable, and ubiquitous, yet increased communication raises patient privacy concerns~\cite{shealth6g}~\cite{privacyfed}. Security and privacy concerns for 6G are being actively considered from the very beginning, with the identification of particular security requirements, highlighting of security challenges, and identification of gaps for future research. Moreover,  emerging technologies such as DLTs and differential privacy are being explored as effective methods to significantly reduce the scale of attacks and personal data breaches~\cite{6gprospective}. Although privacy issues are present, the emphasis on adaptive security in 6G for sustainable healthcare is robust, guaranteeing the protection of patient data and the preservation of privacy.
\section{Conclusion}
The emergence of 6G technology has the potential to transform the healthcare industry. Nevertheless, the increased connectivity and data sharing that comes with 6G also present significant challenges in terms of security and privacy. This article has advocated for the indispensability of an adaptive security architecture in foreseeing the long-term viability of healthcare systems enabled by 6G technology. An architecture of this kind would not alone safeguard sensitive healthcare data but also guarantee the resilience and reliability of healthcare services in the presence of possible cyber hazards. Adaptive security can contribute to enhancing overall resource efficiency by deploying lightweight security solutions upon assessing the network dynamics and security requirements. We anticipate that this discussion will stimulate more investigation and advancement in this crucial field, facilitating the establishment of a secure and sustainable healthcare ecosystem supported by 6G technology. We plan to comprehensively analyse the results
and evaluate the performance of the proposed architecture concerning energy and computation efficiency while
dynamically responding to security \& privacy threats
considering the changing network conditions.

\noindent
\textbf{Acknowledgements.} This research is supported by the Business Finland projects Tomohead (grant 8095/31/2022), Eware-6G (grant 8819/31/2022), Sunset-6G (grant 8682/31/2022), and the Research Council of Finland 6G Flagship program (grant number 346208).
%
%
%
\bibliographystyle{splncs04}
\bibliography{ref}
\end{document}